\documentclass[journal]{vgtc}                

\ifpdf
  \pdfoutput=1\relax                   
  \usepackage{graphicx}                
  \DeclareGraphicsExtensions{.pdf,.png,.jpg,.jpeg} 
\else
  \usepackage{graphicx}                
  \DeclareGraphicsExtensions{.eps}     
\fi%

\graphicspath{{figures/}{pictures/}{images/}{./}} 

\usepackage{microtype}                 
\PassOptionsToPackage{warn}{textcomp}  
\usepackage{textcomp}                  
\usepackage{mathptmx}                  
\usepackage{times}                     
\usepackage{cite}                      
\usepackage{tabu}                      
\usepackage{booktabs}                  

\usepackage[table,xcdraw]{xcolor}
\usepackage{xcolor}
\usepackage{multirow}
\usepackage{subcaption}

\usepackage{fontawesome}
\usepackage[shortlabels]{enumitem}
\usepackage{makecell}
\usepackage{graphicx}
\usepackage{times}
\usepackage{comment}
\usepackage{amsmath}
\usepackage{enumitem}
\usepackage{array}
\newcolumntype{L}[1]{>{\raggedright\let\newline\\\arraybackslash\hspace{0pt}}m{#1}}
\newcolumntype{C}[1]{>{\centering\let\newline\\\arraybackslash\hspace{0pt}}m{#1}}
\newcolumntype{R}[1]{>{\raggedleft\let\newline\\\arraybackslash\hspace{0pt}}m{#1}}
\definecolor{bird}{HTML}{F8766C}
\definecolor{movie}{HTML}{15BFC4}

\usepackage[bookmarks,backref=true,linkcolor=black]{hyperref}
\hypersetup{
  pdfauthor = {},
  pdftitle = {},
  pdfsubject = {},
  pdfkeywords = {},
  colorlinks=true,
  linkcolor= black,
  citecolor= black,
  pageanchor=true,
  urlcolor = black,
  plainpages = false,
  linktocpage
}

\onlineid{1025}

\vgtccategory{Research}
\vgtcpapertype{theory/model}

\title{How Do Data Science Workers \\Communicate Intermediate Results?}

\author{Rock Yuren Pang, Ruotong Wang, Joely Nelson, Leilani Battle}

\authorfooter{
\item
 Rock Yuren Pang is with the University of Washington. E-mail: ypang2@cs.washington.edu.
\item
 Ruotong Wang is with the University of Washington. E-mail: ruotongw@cs.washington.edu
\item
 Joely Nelson is with the University of Washington. E-mail: joelyn@cs.washington.edu 
\item 
 Leilani Battle is with the University of Washington. E-mail: leibatt@cs.washington.edu
}

\shortauthortitle{Pang \MakeLowercase{\textit{et al.}}: How Do Data Science Workers Communicate Intermediate Results?}

\abstract{
 Data science workers increasingly collaborate on large-scale projects before communicating insights to a broader audience in the form of visualization. While prior work has modeled how data science teams, oftentimes with distinct roles and work processes, communicate knowledge to outside stakeholders, we have little knowledge of how data science workers communicate intermediately before delivering the final products. In this work, we contribute a nuanced description of the \textit{intermediate communication} process within data science teams. By analyzing interview data with 8 self-identified data science workers, we characterized the data science intermediate communication process with four factors, including the types of \textit{audience}, \textit{communication goals}, \textit{shared artifacts}, and \textit{mode of communication}. We also identified overarching challenges in the current communication process. We also discussed design implications that might inform better tools that facilitate intermediate communication within data science teams.
} 


\begin{document}


\firstsection{Introduction}
\maketitle

Data science communication often refers to conveying the final analysis insights to a broader audience~\cite{kandel2012enterprise, crisan2020passing, zhang2020data}. For example, researchers and companies increasingly communicate information through high-quality interactive visualizations and dashboards. The New York Times leverages its rich visualization to inform the general public about topics including elections, climate change, and sports. To support more effective final-stage communication, researchers and organizations have developed powerful visualization tools --- such as D3.js, Vega-Lite, Idyll, Tableau, and Microsoft PowerBI --- to simplify the development cycle, condense large-scale dataset, and enable the final information communication with the audience with highly polished visualization. 
 
  However, communication among team members also exists throughout the lifecycle of large-scale data science projects, increasingly in a collaborative fashion~\cite{zhang2020data, mullerdiscover, mullerdiscover, wang2019data}. Today, such collaboration involves multiple different team players and separate analysis stages ranging from data cleaning to visualizing sophisticated findings~\cite{kandel2012enterprise, wang2019data}. Similar to communication in the end, communication during the project also involves explaining technical terms to a non-technical audience (e.g., managers). However, compared to final-stage communication, intermediate communication can focus more on communicating and receiving feedback. To add more complexity, individuals might adopt unique tools at the intermediate stages of a data science project~\cite{zhang2020data}. These resulting intermediate artifacts tend to be far less polished and could be produced by a wider range of tools. We seek to answer the question: \textbf{How do data science workers communicate data intermediately before shipping their final product?}

 We define \emph{intermediate communication} as the synchronous or asynchronous decision-making process where team members build and iterate on the end artifacts for the target audience. In contrast to prior work that categorizes communication as the final step in the data science workflow~\cite{wang2019human, kandel2012enterprise}, we argue that intermediate communication should be a distinct collaboration element where data science workers develop, share, reuse, document, and store analysis with other team members throughout the project lifecycle. Advanced visual analysis authoring tools enable faster data visualization prototyping~\cite{falx, PI2, bauerle2022symphony}, but they largely tackle the engineering side of the problem (i.e., how to make it easier to make interactive visualization). However, data science is engaged as an exploration process more than an engineering process~\cite{zhang2020data, mullerdiscover, heer2008design}. This exploration process can be more flexible and diverse, involving different goals, shared artifacts, and audiences.

 In this paper, we contribute a more nuanced understanding of the intermediate communication process. For example, how teams share resources, resolve conflicts, and seek help.
We conducted eight in-depth interviews with people who self-identify as data scientists/analysts (in industry and academia) and regularly need to communicate and get feedback on their data science work.
 To answer the overarching research question (i.e. how do data science workers communicate data intermediately), we guided our interview with the following questions:

\begin{itemize}[nosep]
    \item Who do data science workers communicate with?
    \item Why do they communicate with others? 
    \item What forms of communication take place in your project?
    \item What challenges, if any, hinder intermediate communication?
\end{itemize}

In the interviews, we focused on participants’ experiences of communicating and receiving feedback, as well as the challenges they encountered in the process. In particular, \emph{we identified four major factors influencing intermediate communication in data science projects}: 

(1) common goals of intermediate communication, (2) types of artifacts that are shared in the communication process, (3) modes of communication (i.e., synchronous versus asynchronous), and (4) common audience configurations observed during intermediate communication. 
\section{Related Work}
\label{sec:related-work}

\subsection{Data Science Workers and Stages} 
\label{sec:data-science}

 Although data science has become a popular term and has gained an increasing amount of attention over recent years~\cite{schroeder_2021}, there does not exist an agreement on the definition of data scientists including their necessary skills and related work tasks~\cite{chatfield2014data}. Prior literature has used data scientists~\cite{crisan2020passing, wang2021much}, data analysts~\cite{kandel2012enterprise, kim2016emerging}, and data science workers~\cite{mullerdiscover} interchangeably. Mueller et al.~\cite{mullerdiscover} posited that data science is a human activity and people involved in data science are therefore data science workers. People who do the work of data science span across multiple job categories and titles, and the definition of data scientists (or data science workers) is likely to become more diverse over time~\cite{mullerdiscover}. In our study, we use the term \emph{data science workers} to refer to anyone whose primary job function deals with or draws meaningful insights from large datasets. 

 In addition to a variety of data science roles, prior research has proposed frameworks that capture the data analysis process over time. In the knowledge discovery and database (KDD) community, the Cross-Industry Standard Processes for Data Mining (CRISP-DM) established five standard phases for data mining based on a prior KDD model~\cite{feyyad1996data}: business understanding, data understanding, modeling, evaluation, and deployment~\cite{wirth2000crisp}. Though capturing a generic data science workflow, this early data science model did not have explicit references to communication. Visualization research studies have examined data science workers with similar phases. In an interview with 35 data analysts, Kandel et. al~\cite{kandel2012enterprise} formalized the data analysis process as consisting of the phases: discover, wrangle, profile, model, and report. Batch and Elmqvist~\cite{batch2017interactive} conducted a contextual inquiry with 8 practitioners and echoed a similar process but found that visualization is optional at best in analysts' work. Based on Kandel's framework, Alspaugh et. al~\cite{alspaugh2018futzing} formed a six-step pipeline by considering the increasing explorative data analysis process in practice. Recently, Crisan et al.~\cite{crisan2020passing} synthesized a comprehensive model from human-computer interaction, visualization, and data science literature with retrospective analysis, resulting in four higher order processes which are preparation, analysis, deployment, and communication. 

 However, prior work model \emph{communication} as a separate stage in the data science workflow, often focusing on the end asynchronous communication to the public, akin to digital journalism. In this work, we argue that communication should be considered throughout a data science project and beyond presentation (explained in \hyperref[related-work:communication]{Section 2.2}). 
 
   \begin{figure}
    \centering
    \includegraphics[width=0.8 \columnwidth]{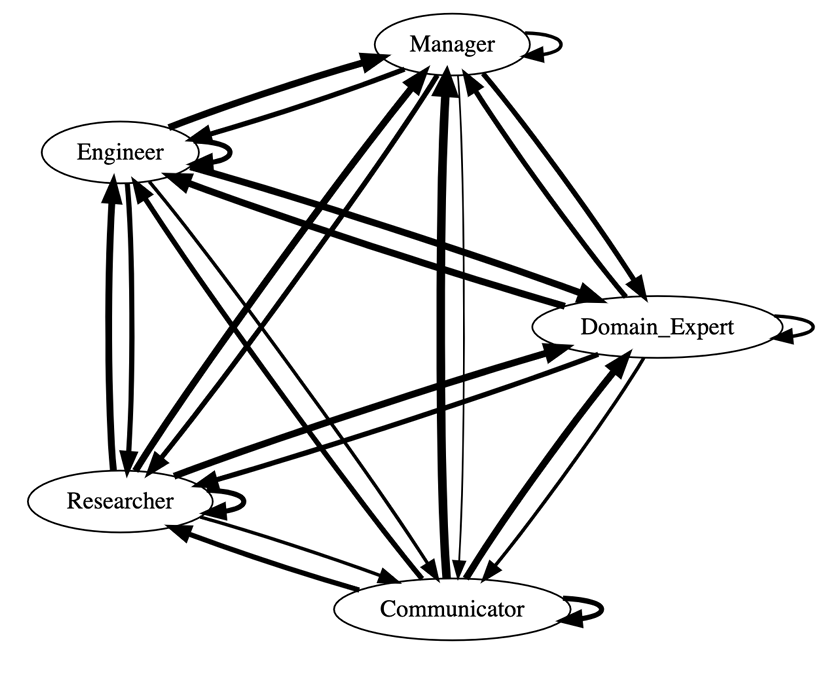}
    \caption{Collaboration among roles in data science projects from a survey of 183 IBM employees. The thickness of each arc represents the normalized proportion of people who reported each directed-type of collaboration. \emph{Communicator} was categorized as a separate role.~\cite{zhang2020data}}
    \label{fig:role}
  \end{figure}
 
 \subsection{Collaboration and Communication in Data Science}
  Recent work recognized that data science is a \textbf{collaborative} process~\cite{miller2014collaborative, borgman2012s} and explored how to support data science stakeholders in the collaboration process~\cite{zhang2020data}. Zhang et al.~\cite{zhang2020data} performed an extensive investigation to understand how data science workers collaborate. In a large-scale survey with 183 employees at IBM, they identified 5 major roles and 6 main stages in collaborative data science workflow as well as tooling and practices for collaboration such as asynchronous data and code documentation. Wang et al.~\cite{wang2019human} identified a "scatter-gather" pattern of collaboration among data science workers. In the "scatter" phase, workers function individually and communicate high-level ideas rather than artifacts in the ensuing "gather" phases. Previous work has also presented in-depth cases study of practices under the contexts of civic data hackathons~\cite{hou2017hacking} and collaboration between domain experts and data scientists~\cite{mao2019data, piorkowski2021ai}. Crisan et al.~\cite{crisan2019uncovering} also identified \emph{collaboration} as an emerging type of data science work within their data science model with an emphasis on data visualization. 
 
  As an important element of collaboration, \textbf{communication} highlights the circulation of artifacts and knowledge. Crisan et al.~\cite{crisan2020passing} categorizes two communication processes: \textbf{documentation} and \textbf{dissemination}. Documentation records and describes data science work and its artifacts, often in the form of capturing the data and analysis provenance~\cite{crisan2020passing, kandel2012enterprise}. In response, recent tools have been developed to address the documentation process, mostly asynchronously capturing the data provenance or documenting analysis processes~\cite{aperitif, davidson2008provenance, ragan2015characterizing}. For example, code gathering tools enabled data analysts to review all archived versions of code outputs and recover the subsets of code that produce the outputs~\cite{managingmess}. 
  URSPRUNG is another transparent provenance collection system designed for data science environments~\cite{rupprecht2020improving}. Tracking the data provenance has also been implemented and tested on computational notebooks to support collaborative data science work~\cite{bestpracticecollaboration, variolite, kery2018story,  wang2019data, rule2018aiding, wang2020callisto, wu2020b2, wang2022diff, yang2021subtle}. Notably, StoryFacets~\cite{StoryFacets} was designed to maintain the visual provenance of the analysis to mitigate barriers to collaboration among audiences  (i.e., expert analysts, managers, and laypersons). 
  
  Dissemination, on the other hand, conveys the insights derived from the data science work, usually as the form of final presentation~\cite{donoho201750, brehmer2022jam}, reports~\cite{wongsuphasawat2019goals} and interactive visualization~\cite{ segel2010narrative}. The emphasis on communicating insights to a larger audience motivated enterprise tools, such as Microsoft PowerBI, Tableau, D3.js, and Vega-Lite, as well as research products that enable rapid visualization development, such as Idyll~\cite{conlen2018idyll}, Falx~\cite{falx}, $\text{PI}_2$~\cite{PI2}, and Symphony~\cite{bauerle2022symphony}. In that regard, tooling largely supports the engineering side of the dissemination process, often formalized as communication as the end step~\cite{crisan2020passing}.
  
  Although collaboration has been categorized as an independent \emph{emerging} type of work~\cite{crisan2020passing}, recent work on collaborative data science also emphasized communicating to the lay audience in the final stage of the pipeline by the roles such as communicator~\cite{zhang2020data}. However, Mueller et al.~\cite{mullerdiscover} suggested that communication might take place throughout the data analysis process. 
  Brehmer and Kosara~\cite{brehmer2022jam} contextualized the intermediate communication by interviewing 23 professionals at Tableau. They identified three scenarios involving presentations of data and provided design suggestions to integrate visualization tools for data analysis and slideware tools for data presentation. More recent work started to address the challenges in intermediate communication. For example, Voder~\cite{8440860} treats data facts as interactive widgets to provide insights throughout the analysis, though they are restricted to system-generated descriptive statistics.
  NB2Slides~\cite{zheng2022telling} generates presentation slides from computational notebooks. Yet, this line of work largely focused on performative presentation which requires a presenter to narrate and step through the content. Our work expands on prior works and covers other aspects of communication such as asynchronous messaging and screen sharing. 

\section{Methods}
\label{sec:methods}
 
 We conducted semi-structured interviews with self-identified data science workers (see \hyperref[sec:data-science]{Section 2.1} for details) to better understand the \emph{intermediate communication} process and needs. 

\subsection{Participants}

 We recruited eight self-identified data science workers (\autoref{tab:pariticpants}) through internal message boards and individual contacts. Of the 8 participants, 4 are current data science professionals, the other 4 are current Ph.D. students but have also been employed at large multinational technology companies. All participants are based in the US. We intentionally recruited participants who have engaged in large data science projects both in academia and industry to holistically extract common themes. Note that our study aims to surface the common goals, practices, and challenges to better formalize in \emph{intermediate communication} in data science projects. In the future, a large-scale survey would be important to quantitatively verify our findings.

    \begin{table}
        \centering
        \begin{tabular}{c|c|c|c}
            \toprule
             \textbf{ID} & \textbf{Role} & \textbf{Exp. (yrs)} & \textbf{Gender}  \\
            P1 & Data Engineer & 2 & M \\
            P2 & Ph.D. Student \& Data Analyst & 2 & F \\
            P3 & Data Analyst & 3 & F \\
            P4 & Ph.D. Student \& Data Scientist & 2 & F \\
            P5 & Ph.D. Student \& Data Scientist & 2 & M \\
            P6 & Data Scientist/Manager & 4 & M \\
            P7 & R\&D Data Scientist & 3 & F \\
            P8 & Ph.D. Student \& Data Scientist & 4 & M \\
            \bottomrule
        \end{tabular}
        \caption{Our 8 participants and their self-reported characteristics: role, years of experience, and gender.}
        \label{tab:pariticpants}
    \end{table}
    
\subsection{Interview Procedures}
 
 We conducted remote, semi-structured interviews with each participant via Zoom or in-person with a recording device for an hour. Our interview started with our project goal and interview instruction. To contextualize the concept of \emph{intermediate communication}, we asked participants to recall their recent data science project where they communicated their findings with other team members. We asked open-ended questions and encouraged interviewees to describe their lived experiences~\cite{kandel2012enterprise}. We organized each interview as follows:
 \begin{itemize}
    \item \textbf{Background} 1) What was a recent data science project you collaborated on with other people? 2) Do you communicate your intermediate results?
     
    \item \textbf{Who do data science workers communicate with?} 3) What was your role in this project? 4) What were the roles and backgrounds of your other team members? 4) What does the overall workflow look like for a project and at what points do you collaborate with others? 
    
    \item \textbf{Why do data science workers communicate with others?} 5) What are your goals for communicating with others? 6) When are you communicating what kind of feedback did you receive from your collaborators? 7) How do you give feedback to your colleagues?
    
    \item \textbf{What forms of communication take place in your project?} 8) How do you communicate with your collaborators (e.g. research meetings, Slack short messages, etc)? 9) What do you often talk about in your communication channel? 10) How did you present the intermediate analysis results with your collaborators? 11) In your data analysis, do your collaborators prefer different tools? If so, how do you resolve the conflicts? 12) How many tools do you use when analyzing your data? 13) How often do you shift between the tools? Why shift? 
    
    \item \textbf{Challenges}: 16) What are the challenges for preparing small group feedback sessions? 17) What are the challenges when communicating your final results?
    
 \end{itemize}
 
 We recorded our interviews and transcribed them verbatim with permission on Marvin by UserFocus \footnote{https://heymarvin.com/}, an interface for coding interviews and sharing these codes across a team.
 
\subsection{Analysis}
    Our analysis followed Braun and Clark's guideline on thematic analysis~\cite{braun2006using}. As all three authors participated in the 8 interviews, we initially engaged with the transcribed text during data collection. We also summarized our takeaways from our interview notes. The authors used open coding to develop initial codes on the Marvin platform based on the transcribed interview texts. We met in multiple sessions to collaboratively revise our codes, ensuring that labeling was consistent across all interviews. We adhered our coding to our original research questions but remained open to observing new themes. We did not calculate the inter-rater reliability, as the primary goal of our study was to surface overarching concepts rather than achieving high agreement among the coders~\cite{mcdonald2019reliability}. 
\section{Results}
   In this section, we present our findings across three subsections, each of which corresponds to one of the top-level themes that emerged through our analysis. 
  
  \subsection{Audience} \label{communication_audience_section}
  
  We observed that data science workers' intermediate communication generally follows the pattern in Figure \ref{fig:role} except the \emph{communicator} role. While not explicitly defined in~\cite{zhang2020data}, the communicator was described similarly to a project coordinator or designer who is responsible for storytelling the final analysis. Our interview surfaced that data science workers regularly initiate communication with other team members throughout the project lifespan. They indicate that communication is an important element in their work, but the audience affects how the intermediate communication. In contrast to distinct data science worker roles~\cite{crisan2020passing, kandel2012enterprise, khan2014big, kim2016emerging}, our study surfaced two main types of audience that affect how data science workers communicate: audience role in the hierarchy and technical expertise. 
  
  \subsubsection{Hierarchy}
 
    Data science workers have different communication styles in mind for the audience in terms of their job hierarchy. 

    \emph{Top-Down} communication occurs when senior data science workers initiate a communication cycle with more junior workers. The communicator may find themselves taking a mentorship role, while the audience has less expertise in technical or project knowledge. P6, a current data scientist manager who mentored two junior data analysts in his previous project, mentioned that he would usually start a conversation to help them "\emph{breaking down problems for them and helping them go through them step by step and help them develop their own intuition}." This would often involve giving them a task and then providing a verbal explanation as to why that task was being done so they would understand [that] it's the purpose and be able to apply it on their own in the future. Notably, several participants mentioned that when they are in a top-down communication cycle, they expected that their audience to understand and implement the tasks but rarely track the progress on their own. As P7 indicates, they suggested ideas to previous interns, only to realize until the next meeting that the intern completely misunderstood them. 
    
    \emph{Bottom-Up} communication occurs when data science workers lead the communication with more senior workers, such as executives, customers, or advisors. The communicator needs to synthesize their complex technical work to understandable and concise insight to their audience in order to show the findings or get feedback. The communication is similar to a performative presentation~\cite{brehmer2022jam} where the audience expects to see the analysis results in a concise manner. To present results, the data science workers create slide decks in an application such as Microsoft PowerPoint or Google Slides. Several participants mentioned presenting their results with the goal to obtain continued support for their project by showcasing only positive improvements of the project. There is less inclination to show positive results in asynchronous communication where participants share screenshots with their managers. As P1 suggested, he felt that it is "\emph{safer and more room to make mistakes}" when messaging his manager via Slack, though he still tries to "\emph{avoid asking questions too many time}".   
    
    \emph{Peer-to-Peer} communication occurred when data science workers actively collaborate with other workers who has a similar role in the hierarchy, with the same ability to determine the project's direction. This peer might have a similar role in the project and a similar level of expertise, although it could be in a different area. This peer may be working on the same task as the data scientist or in a different area on the same project.
    Communication between peers is typically very frank and more casual, compared with bottom-up communication. Peers do not make presentations to share results. Instead, screenshots of the visualization or code would be sufficient to understand the context. They might also casually start a synchronous meeting or screen sharing to solicit feedback on their results, learn about the progress of other members, and troubleshoot problems together.
    
 \subsubsection{Technical Expertise of Audience} 
    The technical expertise of the audience also has an impact on how data science workers communicate their findings. 
    \emph{Non-technical} audience focuses more on the big-picture goals of the work and interpretation of the findings rather than specific details. When communicating with the non-technical audience, data science workers put in considerable effort to make their results understandable to the audience who treat the data processing or modeling process as "\emph{basically magic}" [P6]. When speaking about a project which involved creating a machine learning model to solve a business problem, P6 mentioned that 
    
    \begin{quote}
        "the challenges, at least from my perspective, are trying to explain what was happening to anyone who didn't have a background in machine learning... What you put in your [slide] deck is 'we used this method and you don't even have to name it... They don't know what Bayesian statistics is. They don't know what a Gaussian distribution is. They don't know stochastic gradient descent. So don't say any of that. Say what they will understand." [P6]
    \end{quote}
    
    P6 continued, "\emph{instead of presenting the confusion matrix to them... I would synthesize the confusion matrix into data that [someone without technical expertise] can process without having to explain any math to them}". Similarly, P7 mentioned creating a document for customers "\emph{be more high-level because they might not have the technical understanding to care about the specifics of what was trained.}"
    
    \emph{Technical} audience, however, is able to interpret the concrete data analysis methods. As a result, data science workers often show unedited results either to save time or to receive unbiased feedback about the results. For example, P4 mentioned, "I usually just send [the postdoc] the plot because I don't want to bias them by sharing my interpretation." P7 mentioned having internal documentation for the team going into more detail as to how their machine learning models were trained that would not be shared with the customer unless requested.
    
    \begin{figure*}[h]
        \centering
        \includegraphics[width=\textwidth]{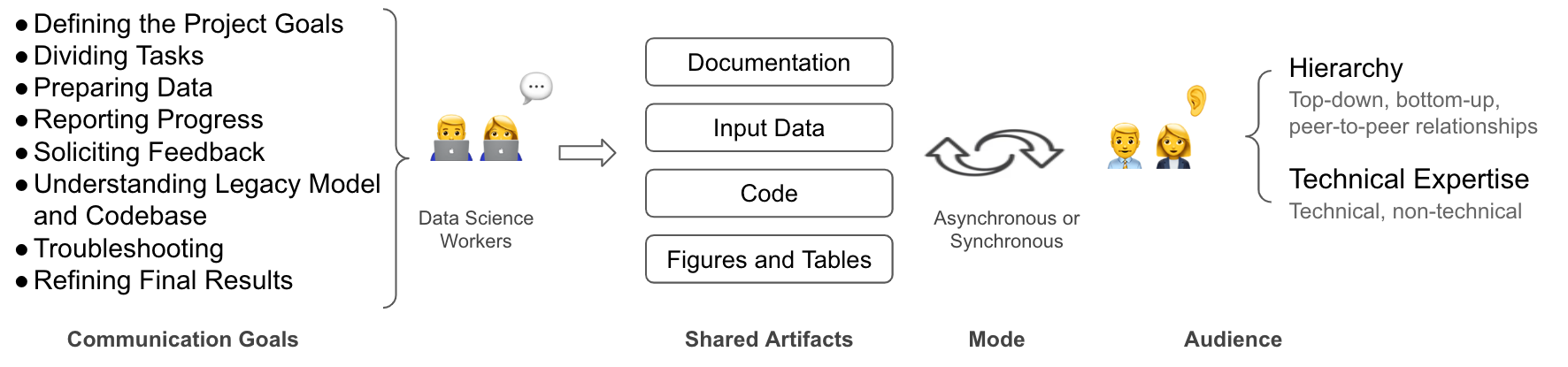}
        \caption{Our model of data science intermediate communication from our interview study.}
        \label{fig:workflow}
    \end{figure*}

\subsection{Communication Goals}
    Data science workers have various goals to communicate with teammates (\autoref{fig:workflow}) instead of working individually. 
    While different communication goals are expected to result in different communication processes and outcomes, we observed workers have mixed goals in one instance of communication. For example, reporting progress and soliciting feedback often take place simultaneously. Additionally, individual data science workers also shift goals within a communication cycle (e.g., prioritize soliciting feedback on an unexpected topic while reporting progress). Therefore, the communication goals in~\autoref{fig:workflow} are not mutually exclusive. Below we detail the eight goals of communication.
   
\subsubsection{Defining the Project Goals} 
    Defining the project goals often start the entire communication cycle or the data science work. All team members, including both technical and non-technical members, collectively establish the high-level research questions, the project scope, and the analysis approach. A feature of the communication process surrounding this goal is that it is frequent and quick. For example, P6's team met at least three times a week with their managers to define the project scope and the client's needs in the first 2-3 weeks. The meeting frequency usually declined to once a week after the team determined the business goal. P8 commented that he, who is a Ph.D. student, would discuss with his advisor to make decisions on data collection before he started any data analysis. The decisions include how much data to collect and how to collect it. Defining the project goals ahead of analysis is in line with prior work which formalizes data science process~\cite{crisan2020passing, crisan2019uncovering, kandel2012enterprise, wirth2000crisp}.
    
    However, data science workers can be motivated to iteratively communicate with their teams throughout the project. This often occurs when the initial project goals started off too ambitious or the unexpected technical analysis blocked the analysis plan. For example, P1, a data engineer in industry, mentioned that they would present to their managers the questions that can be answered by their analysis, and their managers would propose additional questions to ask, which they then integrated into their data analysis plan. P5, a Ph.D. student, also reported defining and updating the overarching project and analysis goals with their advisor throughout the project. Defining, or updating, the project goals, in the middle of the project can be difficult to articulate for junior-level workers, as the entire team has been in sync with a prior plan (P7). Concrete findings, sometimes even a bit of courage, are needed to initiate such conversation. For instance, P8 oftentimes brings descriptive statistics and initial visualization of the datasets to meet with his advisor, hoping to get new inspiration in the weekly meeting. He mentioned, "\emph{sometimes I wasn't just sure how to deal with the data, it is a little under pressure to take that to a meeting.}"  
   
\subsubsection{Dividing Tasks}
    Dividing tasks is a key motivation throughout the project, especially since the team includes multiple data analysts. Based on our interviews, dividing tasks is an essential element for all our participants in the intermediate communication cycle, though it was not explicitly discussed in prior work on the data science process~\cite{crisan2020passing, zhang2020data}. Communication by this goal is usually initiated by managers as all participants from the industry indicated that they were assigned certain main data manipulation tasks. P6, a current data scientist manager who mentored two junior data analysts in his previous project, explicitly mentioned dividing up the tasks is an important element in their daily meetings. He met the analysts regularly to break down the problem into specific coding tasks and checked in to ensure that the analysts did not encounter any roadblocks. Hence, he rarely analyzed the data himself in that project and relied on the daily meeting to understand the progress of the data analysts. However, all Ph.D. students in our sample acknowledged that they need to divide up the tasks on their own, even for their managers (faculty members and collaborators). Common communication channels include short meetings, Slack/Teams messages, or project management tools such as Trello. 
   
\subsubsection{Preparing Data}
    As a unique stage of the data science pipeline, data preparation~\cite{crisan2020passing} --- from identifying the suitable dataset to data wrangling --- is a time-consuming~\cite{mullerdiscover, crisan2020passing}, error-prone~\cite{kery2018story, wang2020better}, yet collaborative~\cite{zhang2020data} process. It is also a common goal for communication. Communication motivated by this goal is more technical, including understanding the data, seeking help, debugging, and clarifying data requirements. As described above, P8 would plot descriptive statistics hoping to gather feedback on the project goal as well as how to narrow down the data requirement. When their team members asked them for help, P6, the data science manager, organized an ad-hoc small meeting to walk through the codebase. Communication is especially necessary if the data or analysis needs to pass to a different data science worker (e.g., if the person in charge of data wrangling is different from the one who builds models). For example, P5, an NLP researcher, described that it took him many efforts in sync with his collaborator who experiments with statistical models. They iterated on data preparation and model building in parallel every time. Whenever he updated the dataset, he also Slack message his collaborator in order to maintain the latest model accordingly. Similarly, P4, also the data cleaner on the team, mentioned that she often communicates with her team members via Slack messages to describe the dataset, check abnormality, and even respond to specific requests (e.g., ``\emph{certain lines of the data seems to be corrupted, could you check row X?}'') Noteworthily, neither P4 nor P5 mentioned that they would test or document their code themselves or by other team members. P5 described an instance where his collaborator failed to incorporate the latest dataset in the model, which resulted in sub-optimal performance of the model.
   
\subsubsection{Reporting Progress}
    In our interview, all data science workers have weekly progress update meetings. Such meetings were commonly designed to keep team members in sync but differed in terms of formality. Some workers need to present their progress (P1, P2, P3, P7), whereas other teams do not have such a requirement (P4, P5, P6, P8). These meetings often invite both non-technical and technical audiences (discussed in Section \ref{communication_audience_section}). Participants often share their progress at a high level rather than delve into the analytical details. Unpolished visualization, such as figures and tables, is used to facilitate progress reporting. However, we observed different strategies to select results from different data scientists. For example, while P8 felt comfortable presenting both significant and insignificant results to his advisor, P1 and P6 reported that they would only report results that are positive or result that are related to a specific question to his managers.
    
\subsubsection{Soliciting Feedback} 
    Soliciting feedback is a known concept that contributes to rapid software development~\cite{abrahamsson2017agile} and project management~\cite{pinto1987critical}. In intermediate communication, data science workers would like to receive constructive feedback on the working results to remove roadblocks and determine the next direction. P1 and P4 mentioned that they appreciate concrete directional feedback such as `\emph{`have you tried this [data transformation]}''. As P1 mentions, this is a sign that the audience (in this case the manager) understands their analyses and roadblocks while he also can act upon the feedback. To solicit feedback successfully, participants report that it is necessary for the data science workers to give enough context, prepare the right questions, and even present intermediate visualization. For example, participants prepare a list of specific questions ahead of each meeting (P2, P4, P5), though coming up with a good question is not always easy. To elicit feedback, our participants also adjust (unpolished) figures to help the audience understand the data and model, though adding necessary contexts such as proper legends, baselines, and axis labels might take too much of their time, especially since these figures are likely to change in the future.

\subsubsection{Understanding Legacy Models and Codebase} 
    Understanding legacy model and codebase is necessary as the data science projects grow larger~\cite{crisan2019uncovering} and new members join the team. For example, P1 commented that he, as the new team member, often needed to understand the existing data infrastructure~\cite{demchenko2012addressing} such as the APIs and input and output of the models in the existing pipeline. He approached this by first reading the internal BitBucket documentation and followed the URLs to configure the environment and understand the data including features and models. He also communicates with the senior team member via Slack messages for clarification. Despite the detailed internal documentation, P1, who worked at two companies, expressed that it is usually much easier and more common to communicate with other more experienced members on the team. Similarly, P7 also commented that she also needed to work with the people who built the model originally, ``\emph{(when retraining the model) I interacted a lot with the person who built the original model to determine what sort of data inputs is required or any nuances [of the data] that needs to be considered. For example, whether the data need to be pre-processed in a certain way.}'' 

\subsubsection{Troubleshooting} 
    All participants in our study indicated that communication to troubleshoot or debug with their peers happens throughout their project. While it can be a discussion in a weekly meeting, troubleshooting often occur on an ad-hoc basis. For example, all participants messaged their colleagues when their code failed rather than waited to resolve them in a weekly meeting. P7 remarked that he would ``\emph{roll my [his] chair over and talk to one of the data analysts}'' when they asked for help. During the pandemic, P6 switched to Discord where he can ``\emph{stream our screen to each other ... [and] immediately see what was happening.}''
    
    Communication to troubleshoot often happens between colleagues who work close to each other. For example, P7 mentioned that the most effective debugging session typically happens with people who know their job well, which are people who work on the same pipeline and not people who are on the same team but work in parallel. Additionally, P3 also mentioned the importance of design for feedback in these sessions. She mentioned that there are regular synchronous code review meetings in her group, where 3-4 attendees from the team participate. These meetings are most effective when people bring not only the code they wrote, but also specific questions that they expect to get answers for. Otherwise, there will be a lot of time wasted in aimlessly browsing through the code.  
 
\subsubsection{Refining Final Results} 
    Towards the end of the project, data science teams communicate to refine the final results and storytelling, oftentimes through interactive visualizations and static visualization on presentation. Communication to refine these visualizations often take place at synchronous meetings, where some team members might bring a draft figure and receive feedback from others. P8 mentioned that he meets with his advisor regularly to iteratively refine figures. While the figures used to facilitate intermediate communication (e.g., figures showing descriptive statistics) focus on including enough information, those on the final presentation are more carefully designed to support a story. The final figures often combine multiple earlier figures and remove irrelevant details to highlight the story.
    
\subsection{Shared Artifacts} \label{artifact}
    Data science workers generate artifacts and leverage these artifacts to facilitate the intermediate communication.

\subsubsection{Documentation} 
    Documentation refers to any decisions, instructions, code or data comments, and provenance in the data science intermediate communication process. Documentation might be stored in a text or markdown file such as Google Doc, Microsoft Word, Github/GitLab, and Bitbucket. Most participants document their analysis decisions and data instructions, oftentimes as a corollary of meeting preparation, though P1 wished to a more explicit way to visualize how decisions change over time. 
    
    On the other hand, provenance tracking of analysis plans, data, code, experiments was not under careful documentation. P5 mentioned that experimental setups can be specified in multiple different places which makes documentation difficult. Documentation on computational notebook pose even more challenge, as P4 and P5 acknowledged that they do not think computational notebooks are intended to share with other team members. Both participants use computation notebooks (in this case Jupyter notebook) for rapid data manipulation and experimentation and intentionally chose not to share with team members or push to the code repository if not asked. 

    Documentation would be easier if a documentation structure has been established. Three participants in industry (P3, P6, P7) have standardized documentation structure. For example, P3 commented ``if you get that good structure at the beginning [of the data science project] then I think it it will work much better down the line and then people won't be confused when they join the project for the first time.'' P7 echoed their team members ``\emph{keep a detailed internal Wiki page on projects}'' and each team members are required to document their experiments and high level takeaways using an existing GitLab issues template. Despite the rigid structure, P7 expressed that she did not document all details and might have missed important findings. She explained ``\emph{We don't document everything. Let's say you run something and it breaks and you try 50 things but then finally something works. We don't document the 49 that don't work. We just document the one that did work.}''

    With a good structure, readable and detailed documentation benefit the audience. P1 mentioned that his team maintains a very thorough documentation, which made his onboarding easy. P3 said that documentation and a readable codebase "\emph{really helps people, especially when they're a new hire. They'll come to the team and be like 'Okay, even if I have no clue what these scripts are doing yet, because I haven't had any chance to look at them, I can still see that there's a very clear order in which things should be run here'}". 

\subsubsection{Input Data}
    Data that is used in analysis and to train models is vital for data science tasks. Often this data is generated by each individual data science worker but is shared among team members. Data is typically stored on a shared server or on a service such as Google Drive. These files are usually in a table format such as an RDS, CSV, or Excel file. Due to the large size of files, they are not usually stored on a version control service such as Git. 

    Communication --- such as data generation, data formatting, and file organization ---- among data science workers is common. P6 mentioned a large amount of collaboration was required to find suitable data sets for their project. The team met several times per week to brainstorm where they could find data sources and communicate their need for data with others around the organization. Data often needs to be in a specified format before it can go into a machine learning model. One participant said they often communicated with their team to "\emph{make sure the data is in the exact format it needs to be in before it goes in between the modeling process}". Having a consistent file structure that all team members could agree on was also an important aspect several participants mentioned. P7 said that at first, her team was experiencing challenges from every person organizing their files and folders differently. The team spent a month discussing the best way to organize their file structures to ensure consistency.
    
\subsubsection{Code} 
    Data scientists often write code in R and Python that they share with others. This code might be stored on a version control system such as Git, a shared server that everyone has access to, or could be shared via other communication channels such as Slack. When troubleshooting, data scientists may share their screens or ssh into a shared machine so others can see their code. Many participants expressed that they had experienced challenges with regard to version control as well as python environments. Keeping code and environments consistent between team members requires considerable effort, especially when many small changes are being made, that they sometimes fall behind on.

\subsubsection{Figures and Tables} 
    Many intermediate results consist of descriptive figures or metrics generated by the analysis. These figures and metrics are useful tools for communicating the current progress of the task and the results of the analysis or model. These are output by the tools and then either exported from that tool or saved by taking a screenshot. One participant noted that "\emph{most of the visualizations are pretty much just static}", and this seemed to be a consistent sentiment across participants. Figures and metrics will be shared in both the intermediate and final stages of the project. Usually, the final figure is much cleaner and more careful decisions are made with regard to the visualization encoding than a figure from the intermediate phase.

\subsection{Mode of Communication: Synchronous and Asynchronous}

\subsubsection{Synchronous Communication} 
    Synchronous, or real-time, communications happen when information is exchanged in real-time, often in the forms of meetings, phone calls, and screen-sharing. In addition to highly organized and formal meetings~\cite{brehmer2022jam}, we highlight the type of \emph{impromptu} synchronous communication exchanges.
    This type of communication features a clear and short task (e.g., how to write the code to apply data transformation). 
    For example, P6 mentioned that their team would always sign into the same remote machine in order to work on the same data and code in real-time. Then, he provides quick instruction to the team members or tackles the problem together. The decisions made in synchronous meetings will then be implemented asynchronously. 

    Our interview surfaced two challenges in synchronous communication: the pressure of peer programming and the difficulty to give feedback on the fly. P8 did not find synchronous work sessions necessarily productive and also found it stressful to code when others were watching. Regardless, they need to do it because current tools, such as Google Collab, do not have real-time updates like Google Docs. You can't experiment with multiple versions but still maintain a shared version. When it comes to generating visualizations, it often takes a while to generate something meaningful which can feel like a waste of time in a synchronous meeting. P7 talked about the challenges of providing synchronous meetings, "\emph{when I see some results I need like 20 minutes to like to think about it. Maybe do some research to determine other interesting areas to pursue. And I think part of it is like when you're just on the spot it's often hard to come up with a good answer.}"

\subsubsection{Asynchronous Communication} 
    Asynchronous Communication happens when messages are not exchanged in real-time. Tools like Slack, MatterMost, or email facilitate this communication. The documentation kept on the experiments and code also serves as a form of asynchronous communication. Since communication in an asynchronous setting may have a large wait time in between messages, workers can have time to produce figures and provide insights without being put on the spot. In addition to challenges to track provenance in computational notebooks~\cite{kery2018story,  wang2019data, rule2018aiding, wang2020callisto, wu2020b2, paths}, asynchronous communications also lack detailed context explanation. One participant found that it was hard to provide the necessary information to have an in-depth conversation in a Slack session, often resulting in ad-hoc synchronous meetings. Another participant expressed that it was easy to get blocked while waiting for a response from a team member.

    Most participants use both modes of communication. Often, feedback provided in synchronous meetings is then implemented and communicated about in asynchronous settings. When a task is taking too long over an asynchronous channel, teams might move to a synchronous channel to resolve issues faster.

\label{sec:results}

\section{Discussion \& Future Work}
\label{sec:discussion}
 
 Our work presents the notion of intermediate communication in data science collaboration as the synchronous or asynchronous decision-making process where team members build and refine the final artifacts for the audience. 
 Despite the common goal to build one end visualization in the visualization community, the process of intermediate communication that leads to the end artifact involves another level of complexity. This work models this process with four different factors that are not fully captured in the prior characterization of the data science work~\cite{crisan2020passing, kandel2012enterprise, zhang2020data, mullerdiscover}. 
 
 Our findings lay the groundwork for thoroughly understanding how various shared artifacts are used, a necessary step before formalizing the specific design requirements for intermediate communication systems. Such a system is crucial as our study shows the complicated picture as different audiences can have different goals. This is different from general communication in that this communication surrounds unique data science shared artifacts. General-purpose artifacts, such as a presentation or a Google Doc, also can be associated with the intermediate results. These intermediate artifacts all have an impact on the end visualization product. In our analysis, we made an initial attempt to surface the current artifact usage in the intermediate communication model, which we use three examples to demonstrate as below. 
    
    \textit{Sharing static figures and tables for interactive goals}: In our interview, all participants indicated that they would share a screenshot of a visualization or analysis table when communicating with colleagues. P2, a Ph.D. student, mentioned that she often shared a screenshot of a static visualization. Specifically, she reported that her most recent experience involves creating scatter plots of millions of data points. During a research meeting, her advisor would always like to check the information of a specific point on the plot. Even though she is aware of interactive data visualization tools such as D3 and Vega-Lite, she feels that it would be adding significant mental and engineering efforts to build a usable interactive application. She said that her goal was not to create only one scatter plot. Instead, she had to constantly change the data frame and the visualization to explore the data. As a result, changing toolkits is simply too much work for her. Similarly, all participants in our interview seem to agree that their focus is on their assigned task (e.g., data cleaning, model building, etc.), making their intermediate visualization interactive would greatly improve the quality of their meeting. However, it is "\emph{just not the first thing on my mind}", as P6 said. In this example, reporting progress and soliciting feedback might require interactive visualization to rapidly detect outliers of interest. However, mapping static visualization --- which is usually an end result by the data science worker --- to an interactive visualization requires many engineering efforts that are not central to the worker's task. In other words, the shared artifacts, currently as they are, do sufficiently support the data science collaborative environment which requires interactive interaction.
    
    \textit{Sharing different artifacts with different audience}s: Our interviews showed that participants shared different artifacts depending on their audience. In bottom-up communication, workers would synthesize analyses and only showcase minimal analysis to stakeholders, and executives in a form of presentation, whereas peer-to-peer communication, is more holistic. All participants wished to communicate all their intermediate analyses as a bottom-up communicator but only selected ones that showed significant insights. P1 said, "\emph{I wish my boss would look at all my analyses to make sure that I don't miss any important findings, but sometimes I might look silly if I throw everything at him.}" Besides communicating in meetings, P5 said that "\emph{my teammates know my analysis to avoid duplicate work, but the code on Github doesn't usually reflect that.}" In this example, bottom-up communication requires selected analysis results in the interest of the audience's time. However, the intentional omission of insignificant results pose threat to the artifact validity due to the reporting bias. In other words, the act of preparing communication artifacts may reveal (or even induce) threats to validity in the underlying analysis. Communication, especially intermediate communication, is not just reporting results; communication is inextricably linked with the analysis itself. Communication sensemaking likely influences analytic strategy, just as analytic sensemaking influences communication strategy. Thus, intermediate communication artifacts should be evaluated with a similarly critical lens.

    \textit{Sharing multiple artifacts asynchronously over the data science lifecycle}: Asynchronous communication requires sharing data science artifacts, including input data, analysis code, visualizations, and documentation. However, data scientists need to navigate different tools to maintain the consistency between artifacts. In our interview and the authors’ experiences, input data is usually shared on an online server (e.g., an internal server, Google Drive, etc.). Sometimes, data scientists informally share an intermediate data file over communication channels (e.g., Slack and Email), thus making it difficult to trace the data file version. Data scientists share code and computational notebooks on version control systems (e.g., Git) and online servers. Documentation is maintained on communication channels (e.g., Google Docs and Slack) and code hosting platforms (e.g., README file on Github). However, our interviews revealed that code, visualizations, and documentation are not easily stored by our participants, though all accepted that recording this information would benefit their analysis workflow. In this example, asynchronous communication requires data science workers to have access to multiple artifacts. However, navigating different tools and connecting artifacts in the data lifecycle can be a daunting task in practice. The analyst will not want to show organizational leadership the wrong version of an intermediate artifact that contains errors, bugs, confusing visualizations/code, etc. But they also don't want to duplicate their efforts by creating many parallel versions of those same artifacts just for presentation purposes.
 
 The three scenarios exemplify how currently the shared artifacts (Section \ref{artifact}) are used in intermediate communication, but also surface potential gaps in the design of current artifacts and users needs. We noticed that participants currently repurposed existing general-purpose communication and collaboration artifacts (e.g., static figure, Google Doc) to support intermediate data science communication. However, these artifacts could fall short in addressing the unique challenges brought by the complexity and variability of data science collaboration scenarios. This mismatch between the design of current artifacts and users' needs points to the need of developing a more comprehensive \textit{framework} that can help researchers to more systematically reason about (1) the characteristics of successful artifacts for intermediate data science communication and (2) how artifacts can affect intermediate data science communication. In this framework, people and the environment influence intermediate results, but intermediate results can also influence people and the environment. It becomes a spiral that changes with every iteration as artifacts are refined and people are more aware of the data science problem. 
 
 With the framework, data science workers and researchers will be able to more systematically reason about the impact of artifacts in data science communication, which could guide more responsible uses of artifacts to facilitate communication, as well as the better design of future artifacts to support intermediate communication in data science teams. For example, it could tackle the three examples above where current artifacts do not sufficiently support intermediate communication. It could also help developers and researchers understand and trace potential threats to the validity of the artifact.
 
 Our research is an initial step towards such a framework, but comprehensively building it warrants future work. Our work identified the unique four factors that characterize the intermediate communication process in data science teams, but future work needs to systematically categorize and understand the artifact usages for different goals, among different audiences, and across different modes. Such a framework will guide an ultimate intermediate communication tooling support. Below, we lay out the research plan for building a framework. An immediate next step is collecting more empirical evidence on how artifacts are used in the intermediate communication process in data science collaborations. Our work described the different ways that data science workers communicate in detail and hinted at the opportunities for better artifacts to support such communication, but we were not able to articulate the role that artifacts play in the process. Future work can take on a more tool-oriented perspective to study how people use the artifacts to support different scenarios in intermediate communication. A further next step can also test out how different variations of the design of the artifact (e.g., ease of creation, flexibility, information density) affect the intermediate communication process differently. There may be interactions between each factor, which were not studied in depth in this work. A future framework could capture explicit relationships between the major dimensions studied in this paper.
 
 Lastly, we note that our participants' sample is skewed towards junior data scientists. All participants only have 2-4 years of experience. In communication scenarios, junior and senior team members tend to play different roles. For example, while junior data science workers would prefer to receive clear instructions on the next steps, senior members might not feel comfortable engaging with technical details within the limited prior synchronous meeting time. In our interviews, we noticed some early signs of expectation mismatch between members of different seniority on teams, but our observation was limited by our skewed data. It would thus be helpful to expand our study to a more diverse set of participants, spanning different seniority on teams, as well as covering different types of scenarios. We can also imagine deploying a survey to validate our findings on a broader population.
 
\section{Conclusion}
 
 In this paper, we present analysis results based on interviews with eight data science workers. We identified four factors that affect the data science intermediate communication process, which are \textit{goals of communication}, \textit{artifacts being shared}, \textit{mode of communication}, and \textit{audience of communication}. We also identified an important future direction to build a framework that comprehensively captures the interaction between four factors in our analysis. We hope that our work can inspire future work to expand on this framework and spark new conversation on tooling design that supports the unique intermediate communication in the data science team.

\bibliography{paper}
\bibliographystyle{abbrv-doi}

\end{document}